\documentclass[prl,twocolumn,superscriptaddress,floatfix,noshowpacs,10pt,longbibliography]{revtex4-2}
\usepackage{graphicx,bm,times}
\usepackage{amsmath}
\usepackage{amsfonts}
\usepackage{amssymb}
\usepackage[version=4]{mhchem} 
\usepackage{lipsum}
\usepackage{bm}
\usepackage{color}
\usepackage{times}
\usepackage{xcolor}
\usepackage[%
  colorlinks=true,
  urlcolor=blue,
  linkcolor=blue,
  citecolor=blue
  ]{hyperref}
\DeclareUnicodeCharacter{03B1}{\alph}

\begin{document}

\title{Anisotropy of the zigzag order
in the Kitaev honeycomb magnet $\alpha$-RuBr$_3$  }

\author{John~S.~Pearce}
\affiliation{Clarendon Laboratory, Department of Physics,
	University of Oxford, Parks Road, Oxford OX1 3PU, UK}

 \author{David~A.~S.~Kaib}
\affiliation{Institut für Theoretische Physik, Goethe-Universität Frankfurt, 60438, Frankfurt am Main, Germany}

\author{Zeyu~Ma}
\affiliation{Clarendon Laboratory, Department of Physics,
	University of Oxford, Parks Road, Oxford OX1 3PU, UK}

\author{Danrui~Ni}
\affiliation{Department of Chemistry, Princeton University,
Princeton, NJ 08544, USA}

\author{R.~J.~Cava}
\affiliation{Department of Chemistry, Princeton University,
Princeton, NJ 08544, USA}

\author{Roser~Valentí}
\affiliation{Institut für Theoretische Physik, Goethe-Universität Frankfurt, 60438, Frankfurt am Main, Germany}

\author{Radu~Coldea}
\affiliation{Clarendon Laboratory, Department of Physics,
University of Oxford, Parks Road, Oxford OX1 3PU, UK}

\author{Amalia~I.~Coldea}
\email[corresponding author:]{amalia.coldea@physics.ox.ac.uk}
\affiliation{Clarendon Laboratory, Department of Physics,
University of Oxford, Parks Road, Oxford OX1 3PU, UK}

\date{\today}

\begin{abstract}
Kitaev materials often order magnetically at low temperatures
due to the presence of non-Kitaev interactions.
Torque magnetometry is a very sensitive technique for probing the magnetic anisotropy, which is critical in understanding the magnetic ground state.
In this work, we report detailed single-crystal torque measurements in the proposed Kitaev candidate honeycomb magnet $\alpha$-RuBr$_3$, which displays zigzag order below 34~K. Based on angular-dependent torque studies in magnetic fields up to 16~T rotated in the plane normal to the honeycomb layers, we find an easy-plane anisotropy with a temperature dependence of the torque amplitude following closely the behaviour of the powder magnetic susceptibility. The torque for field rotated in the honeycomb plane has a clear six-fold periodicity with a saw-tooth shape, reflecting the three-fold symmetry of the crystal structure and stabilization of different zigzag domains depending on the field orientation, with a torque amplitude that follows an order parameter form inside the zigzag phase.
By comparing experimental data with theoretical calculations we identify the relevant anisotropic interactions and the role of the competition between different zigzag domains in this candidate Kitaev material.
\end{abstract}

\maketitle

\begin{figure}[h]
	\centering
\includegraphics[width=\linewidth,clip=true]{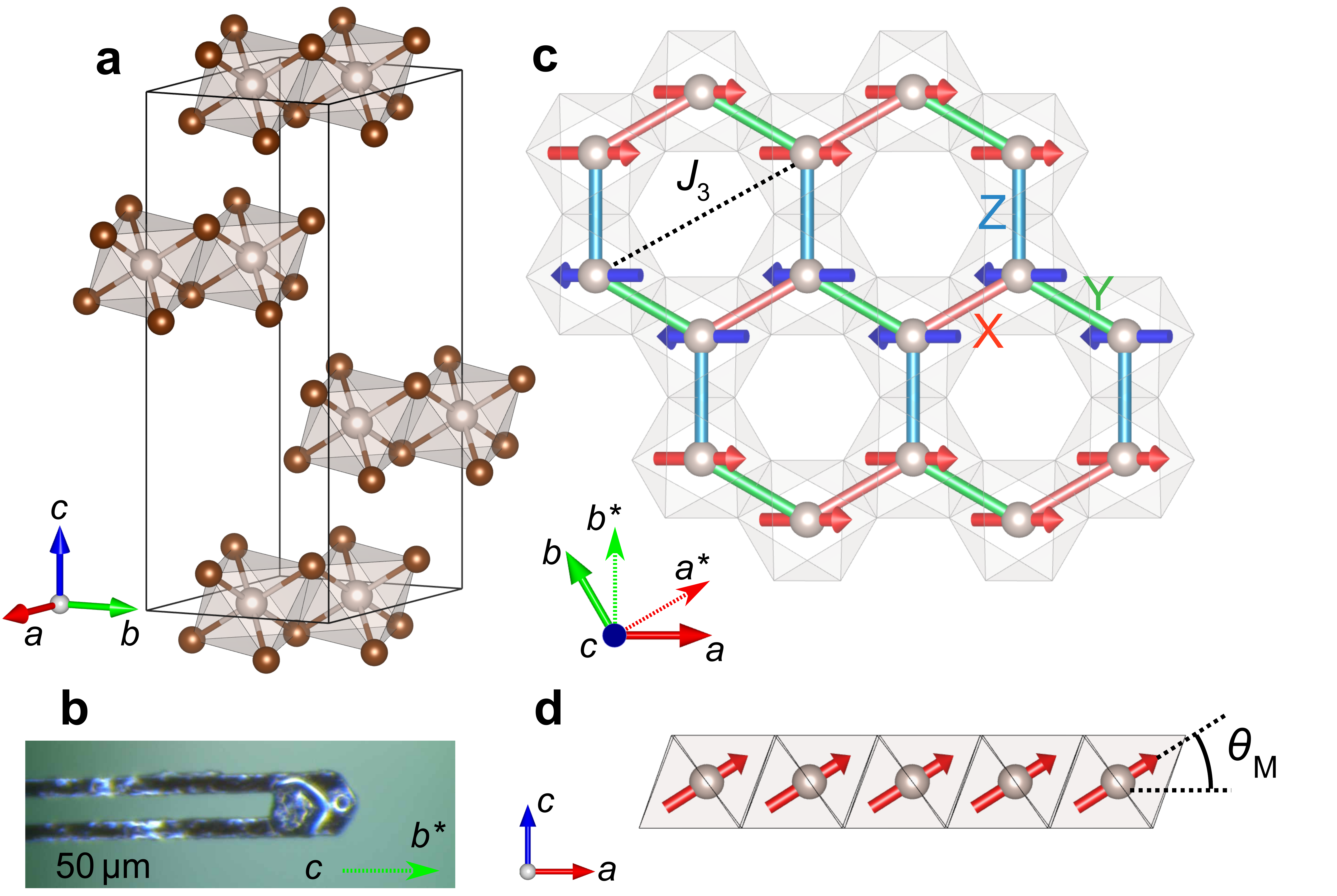}
\vspace{-0.2cm}
\caption{{\bf Magnetic structure of $\alpha$-RuBr$_3$.}
(a) Crystallographic unit cell (space group $R\bar3$ in hexagonal setting) where local Ru$^{3+}$ moments reside inside a network of edge-sharing RuBr$_6$ octahedra. (b) Sample S1 placed on a piezocantilever for torque measurements in the $b^\ast\!c$ plane.
(c) Collinear antiferromagnetic zigzag order in a honeycomb layer
showing oppositely aligned moments (red and blue arrows) canted away from the honeycomb plane by an angle $\theta_{\mathrm{M}}$, as defined  in (d). Diagram in (c) represent a magnetic domain with propagation vector (0,1/2,0); symmetry-equivalent domains are obtained by $\pm120^\circ$ rotation around the $c$-axis. Dashed lines in top left hexagon indicate the exchange paths for $J_3$, essential to stabilize this order.
Colour-coding of bonds indicates the XYZ Ising exchange character for an ideal Kitaev model.
(d) Canting of the moments out of the honeycomb plane
by an angle $\theta_{\mathrm{M}}$ (reported to be in the range $32^\circ$ \cite{Kaib2022} to 64$^\circ$ \cite{Imai2022}) for the magnetic structure in (c).
}
\label{Fig1}
\end{figure}

{\textit{Introduction.}}
 The Kitaev model consisting of  $J_{\rm eff}=\frac{1}{2}$ effective moments interacting via bond-dependent Ising exchange in a two-dimensional honeycomb lattice~\cite{Kitaev2006}  has garnered significant attention in quantum magnetism~\cite{jackeli2009MottInsulatorsStrong,winter2017ModelsMaterialsGeneralized,Takagi2019,trebst2022KitaevMaterials} due to its unique properties of an exactly solvable quantum spin liquid ground state and fractionalized excitations.
The RuX$_3$ family (X = Cl, Br, I),
consisting of  honeycomb lattices of edge-sharing RuX$_6$ octahedra  [see Fig.~\ref{Fig1}(a) and (c)], has been proposed as an ideal family for exploring the relevance of Kitaev interactions~\cite{Ni2022,Imai2022,Kaib2022,Choi2022,Ni2024,shen2024MagneticVsNonmagnetic}.
The ruthenium-based trihalide, $\alpha$-RuCl$_3$,
is a Mott insulator with dominant ferromagnetic Kitaev interactions,
however, finite
Heisenberg and off-diagonal terms are responsible for the presence of antiferromagnetic zigzag order below $\approx7$~K. 
Replacing Cl with a heavier halogen, such as Br, was suggested as a way to
approach the Kitaev limit since the size of Br weakens the direct Ru $d-d$ hybridization \cite{Imai2022} and suppresses non-Kitaev exchange terms (such as Heisenberg and off-diagonal symmetric exchange).
However, it was found experimentally that
the zigzag order in  $\alpha$-RuBr$_3$ (illustrated in Fig.~\ref{Fig1}(c)) is even more stable than in $\alpha$-RuCl$_3$ with $T_\text{N}\approx34$~K~\cite{Imai2022}, as confirmed by muon spin rotation \cite{Weinhold2024} and Raman spectroscopy studies, which indicated enhanced $p$-$d$ hybridization ~\cite{Choi2022}.

Theoretically, a first principles-based analysis of the
exchange interactions in $\alpha$-RuBr$_3$~\cite{Kaib2022} suggested
that the presence of non-Kitaev  exchange interactions due to
enhanced $p$-$d$ hybridization and a complex interplay of magnetic ion and ligand spin-orbit coupling effects promote long-range zigzag magnetic order.
Furthermore, it was found that moderate pressures increase the stability of the zigzag order and this was attributed to enhanced third-nearest-neighbour Heisenberg interactions and interlayer couplings
\cite{shen2024MagneticVsNonmagnetic}.
Despite these advances, a complete picture on the nature of the exchange interactions that stabilize the magnetic order in Kitaev candidate systems
is still missing.

\begin{figure}[hbtp]
	\centering
\includegraphics[width=\linewidth, clip=true]{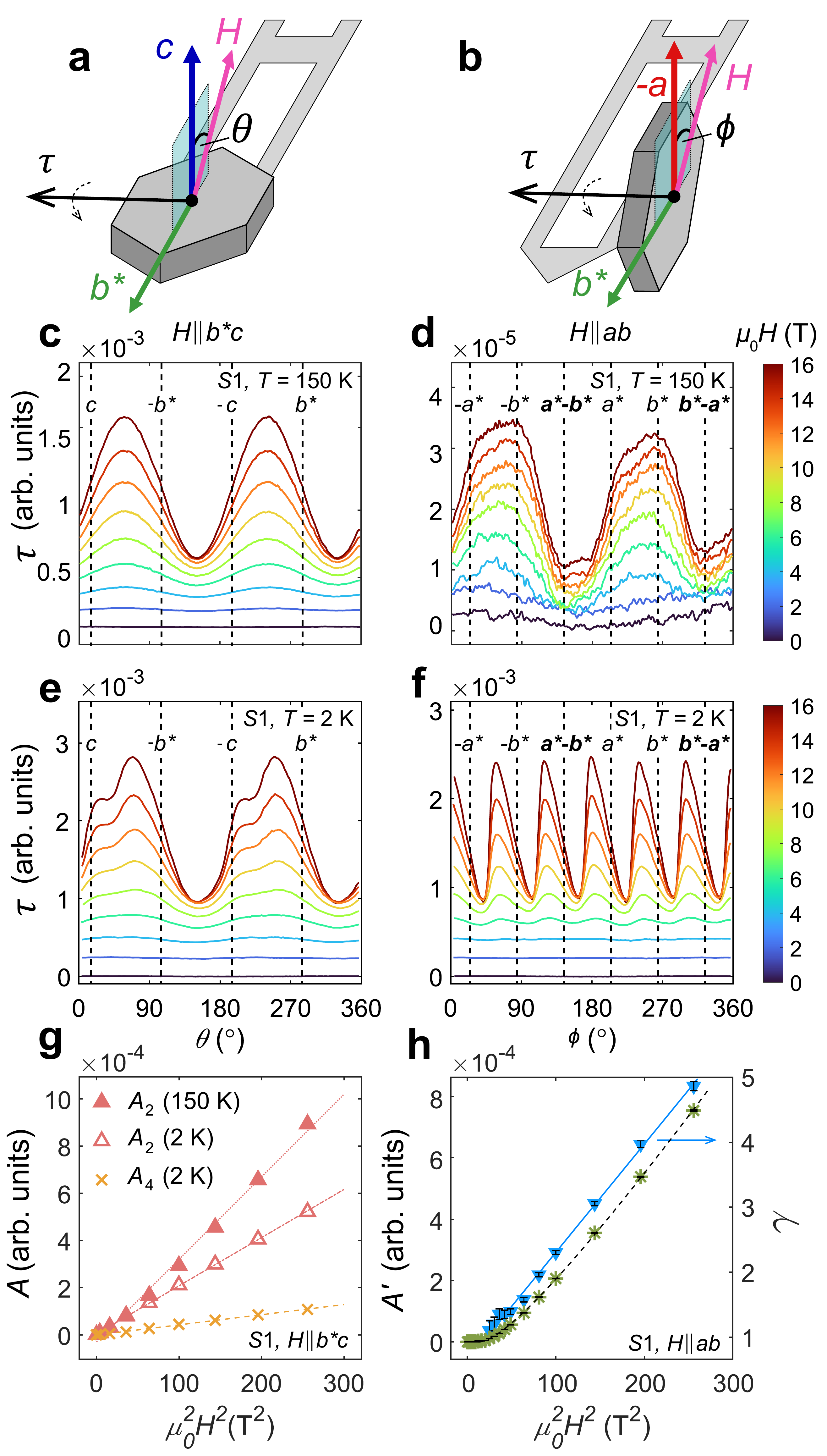}
\vspace{-0.2cm}
		\caption{
  {\bf Angular dependence of the magnetic torque in fields up to 16~T.}
  Diagram of sample S1 mounted onto the cantilever in $H\parallel b^\ast\!c$ and $H\parallel ab$ orientations are shown in (a) and (b), respectively. $\theta$ and $\phi$ are the out-of-plane and in-plane angles, respectively, between the cantilever normal and the fixed applied field direction.
  The angular dependence of the torque for sample S1 in varying magnetic
  fields up to 16~T is shown in (c-f). Measurements were carried out at 150~K (c-d) and 2~K (e-f), with left and right panels corresponding to $H\parallel b^\ast\!c$ and $H\parallel ab$, respectively.
   Curves are offset vertically for clarity.
     Vertical dashed lines indicate where the magnetic field is close to certain crystallographic axes.
   Fourier amplitudes as a function of $H^2$ are plotted in (g) for the $H\parallel b^\ast\!c$ orientation. Two-fold amplitudes ($A_2$) at 2~K and 150~K are represented by filled and open triangles, respectively, while four-fold amplitudes ($A_4$) are represented by crosses. Dashed lines are linear fits to the data points.
  (h) Phenomenological parameters of the sawtooth model,
    $A'$ (stars) and $\gamma$ (filled triangles), plotted against $H^2$.
       The dashed black line is a fit to the form $A' = A'_0(H-H_0)^2$, where $H_0=3.5$~T is attributed to a threshold field to overcome domain pinning effects.
   }
  \label{Fig2}
\end{figure}

To address this, the recent synthesis of crystals of $\alpha$-RuBr$_3$ \cite{Ni2024}
has opened up new opportunities to study its magnetic behaviour
and probe the anisotropic interactions.
In this work we report a detailed torque magnetometry study, as a function of temperature and field strength and orientation in two orthogonal crystallographic planes, on micron-sized $\alpha$-RuBr$_3$ crystals.
Field rotation in the plane normal to the honeycomb layers {[Fig.~\ref{Fig2}(a)]}
identifies that the largest magnetic susceptibility occurs for field parallel to the honeycomb plane, hosting dominant ferromagnetic interactions with a torque amplitude that follows the temperature dependence of the magnetic susceptibility. On the other hand, field rotation in the honeycomb layers {[Fig.~\ref{Fig2}(b)]} reveals a clear six-fold sawtooth torque signal, with an amplitude that has an order parameter behaviour below $T_\text{N}$. By comparing to theoretical calculations on \textit{ab initio}-based extended Kitaev models, we identify this six-fold torque signal to originate from the selection of different zigzag magnetic domains
while the sample is rotating in magnetic field.
Furthermore,
the observed easy-plane anisotropy is attributed
to off-diagonal
exchange terms as well as the $g$-tensor anisotropy.

\begin{figure*}[hbtp]
\centering
\includegraphics[width=1\linewidth,clip=true]{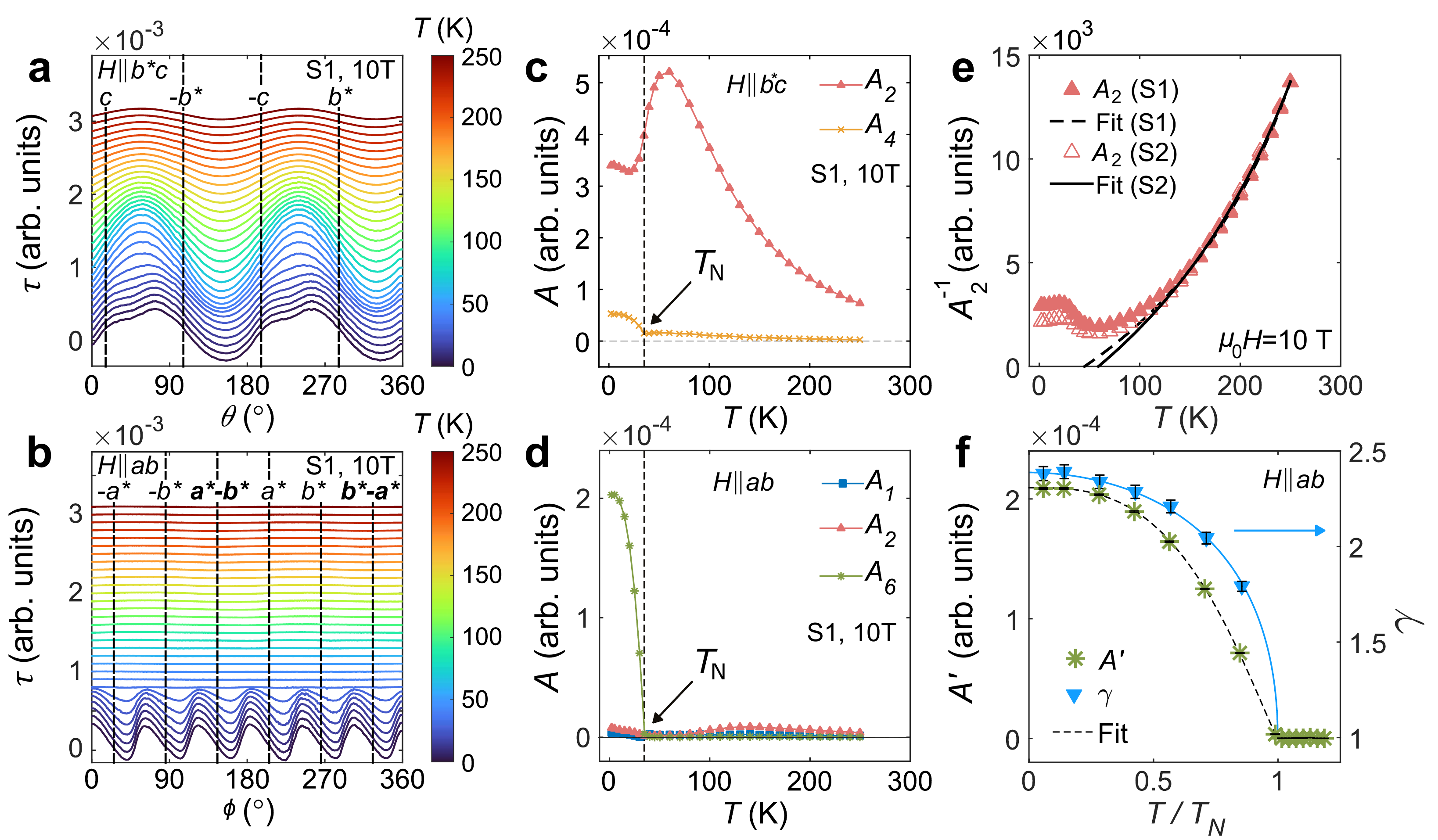}
\vspace{-0.2cm}
  \caption{\textbf{Temperature-dependent magnetic torque studies.}
  (a,b) Angular dependence of the magnetic torque for sample S1
  measured at constant temperatures between 2~K and 250~K in a constant field of 10~T
  corresponding to $H\parallel b^\ast\!c$ in (a) and $H\parallel ab$ in (b). Constant-temperature curves have been shifted vertically for clarity, with the bottom (top) curve in each panel representing data at 2~K (250~K).   (c,d) Fourier amplitudes plotted against temperature, extracted from the data shown in panels (a) and (b). $A_2$, $A_4$ and $A_6$ are represented by triangles, crosses and stars, respectively. Arrows indicate the sudden rise of $A_4$ and $A_6$ at $T_\text{N}$, the zigzag transition temperature. (e) Curie-Weiss fits of the $H\parallel b^\ast\!c$ inverse two-fold amplitude, $1/A_2$, for samples S1 (filled triangles) and S2 (open triangles), normalized at 250~K. (f) Temperature dependence of the phenomenological parameters $A'$ (stars) and $\gamma$ (triangles), obtained from fitting a sawtooth model to the data in panel (b) below $T_\text{N}$. The dashed black line is an order parameter fit to $A'$ as described in the text. The solid line are guides to the eye to $\gamma$  (right axis).
}
  	\label{Fig3}
\end{figure*}

{\it Experimental details.}
$\alpha$-RuBr$_3$ crystals of diameter less than 50~$\mu$m were synthesized using a high-pressure technique \cite{Ni2024}. Samples were screened via single-crystal X-ray diffraction, which confirmed the BiI$_3$-type structure \cite{Imai2022}, with a R$\bar{3}$ space group (no.\ 148)
[see Fig. S2 in the SM \cite{SM}].
Angle-dependent torque measurements were performed on two high-quality samples (S1 and S2) mounted onto piezocantilevers on a cryogenic rotator and cooled down to 2~K using a 16~T Quantum Design PPMS. The sample platforms were rotated such as to probe the torque for magnetic field in two orthogonal crystallographic planes. Constant-temperature studies were carried out at 2~K and 150~K while constant-field studies were performed at 10~T.

{\it Results.}
The angular dependence of the measured torque $\tau$ for magnetic field rotated in a plane normal ($H\parallel b^{*}c$) and parallel to the honeycomb plane ($H\parallel ab$) is summarised in Figs.~\ref{Fig2} and \ref{Fig3}. The angular dependence is parameterised by the Fourier decomposition
$$\tau(\theta) = A_1 \cos{(\theta-\theta_1)} + \sum_{n=2,4,6} A_{n} \sin{[n(\theta-\theta_{n})]},$$
where the first term accounts for the sample weight and the following terms parameterize the magnetic torque. Here $\theta_i$ are angular offsets that account for a small misalignment. $\theta$ is the rotation angle between the cantilever normal and the fixed field direction; the rotation angle is denoted by $\phi$ for magnetic field rotated within the $ab$-plane [see Figs.~\ref{Fig2}(a-b)]. The amplitudes $A_i$ are determined via fast Fourier transform (FFT).

{\it Field rotation in a plane normal to the honeycomb layers} [see Fig.~\ref{Fig2}(a)].
Neutron powder diffraction studies  on powders found that  $\alpha$-RuBr$_3$ develops zigzag magnetic order
below $T_\text{N} \approx34$~K \cite{Imai2022}, as illustrated in Fig.~\ref{Fig1}(c).
In the paramagnetic phase above $T_\text{N}$, we observe a torque signal of the form $\sin{2\theta}$ for magnetic field rotated in the $b^\ast\!c$-plane, as evident in Fig.~\ref{Fig2}(c).
The magnetic torque gives the angular derivative of the free-energy $\tau=-\partial F/\partial \theta$ such that positions of $\tau=0$ with negative gradient, $\partial\tau/\partial\theta<0$ reflect stable equilibria close to in-plane directions
[see Fig. 2(c) and Fig. S6(a) in the SM \cite{SM}].
This infers that $\alpha$-RuBr$_3$ possesses an easy-plane anisotropy, similar to  $\alpha$-RuCl$_3$ \cite{Yang2023, LiWei2021}.

The two-fold torque component, $A_2$, is expected to be proportional to the magnitude of the susceptibility anisotropy,
in this case $\tau \sim (\chi_{ab}-\chi_c)$. The field dependence of $A_2$
at 2~K and 150~K is well described by a parabolic $H^2$ form above and below $T_\text{N}$, as one would expect from an induced magnetization proportional to the applied field both in the paramagnetic and ordered phases, as shown in Fig.~\ref{Fig2}(g). Interestingly, the temperature dependence of $A_2$ [see Fig.~\ref{Fig3}(c)] bears strong resemblance to the temperature-dependence of the powder-averaged susceptibility data
(see Fig.~S10 in the SM \cite{SM}),
where $\chi_{\mathrm{avg}}$ exhibits a broad peak at around 60~K, attributed to the development of antiferromagnetic correlations \cite{Imai2022},
followed by a significant drop with maximal slope near $T_\text{N}$ \cite{Imai2022, Kaib2022}. A Curie-Weiss fit of the $A_2$ amplitude
in Fig.~\ref{Fig3}(e) gives $\Theta_{\mathrm{CW}}=+45(4)~\mathrm{K}$ and $+59(6)$~K for sample S1 and S2, respectively, indicating dominant ferromagnetic interactions in $\alpha$-RuBr$_3$,
in agreement with theoretical predictions \cite{Kaib2022}.

On the other hand, in the zigzag phase below $T_\text{N}$, the angular dependence of $\tau$ develops a camel hump-like feature, as shown in Fig.~\ref{Fig3}(a) and  Fig.~\ref{Fig2}(e).
This feature can be parameterised by the sudden appearance of a four-fold FFT component, $A_4$, below $T_N$ [see Fig.~\ref{Fig3}(c)]
and it also follows a parabolic dependence in magnetic field
[see Fig.~\ref{Fig2}(g)]. Theoretically, a four-fold component for the out-of-plane torque was assigned to a weak cubic anisotropy via a quantum order-by-disorder mechanism in $\alpha$-RuCl$_3$ \cite{Riedl2019}.
However,
a double-hump feature
in sample S1
is only seen in the peaks of the torque ($\theta\sim 30, 210^{\circ}$) and not the troughs, even up to 16~T as shown in Fig.~\ref{Fig2}(e).
Meanwhile,  this feature for sample S2 occurs in the troughs instead
[see Fig.~S6 in the SM \cite{SM}]
and cooling
in magnetic field
aligned along the honeycomb plane did not affect its size or shape
(see Fig.~S7 in the SM \cite{SM}).

\begin{figure}[hbtp]
\centering
\includegraphics[width=1\linewidth,clip=true]{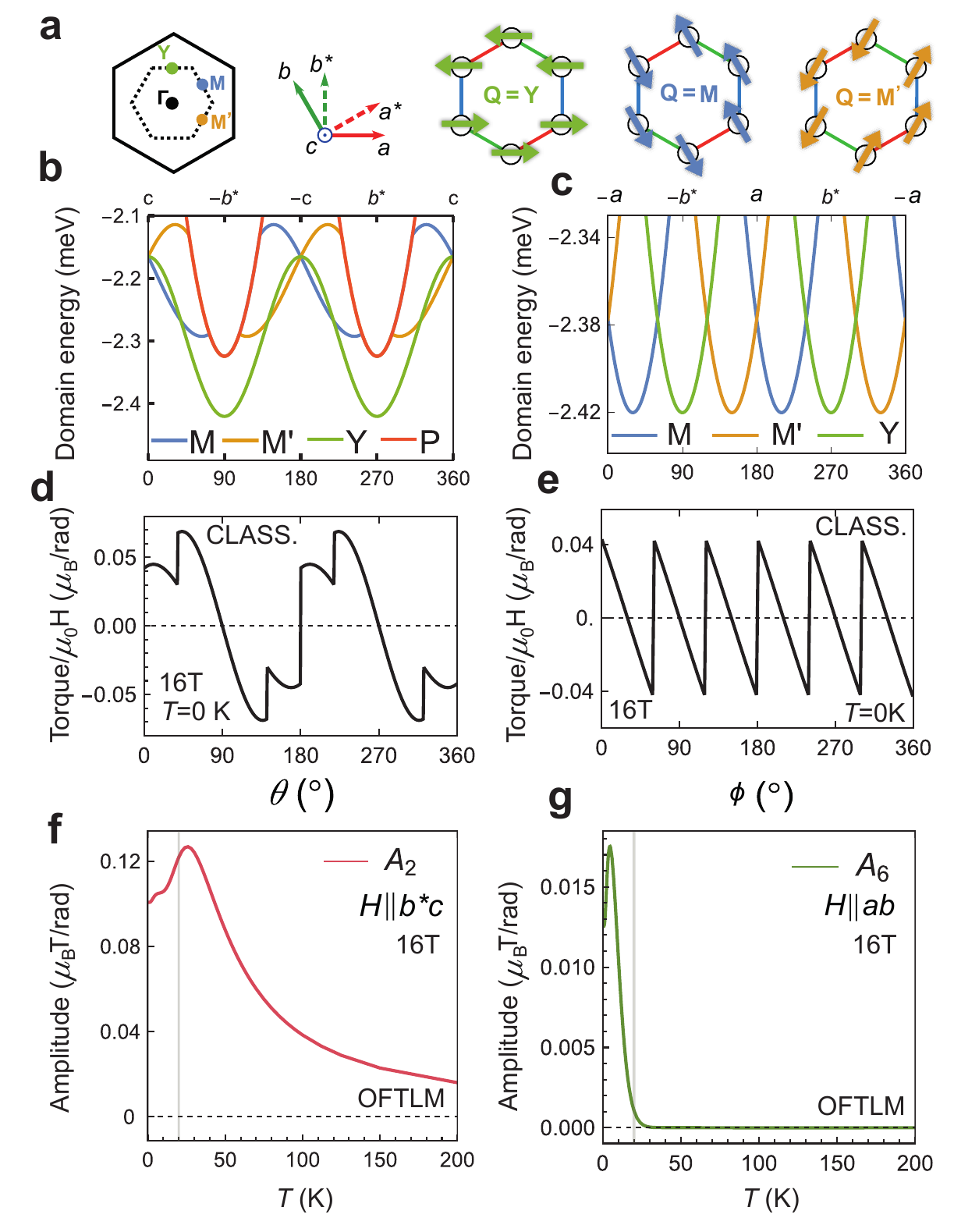}
  \caption{\textbf{Theoretical modelling of torque in $\alpha$-RuBr$_3$.}
       (a) Brillouin zone and three (zero-field) zigzag domains. Solid (dashed) hexagon at the left: Wave vectors within the third (first) Brillouin zone. Y, M and M' are the zigzag wave vectors.
   (b-c)
  Angular dependence of the energy for different zigzag domains (allowing canting of the moments) considering the classical zero-temperature approximation for the two different rotation planes. The red line in (b) corresponds to the field-polarized state.  (d,e) The resulting torque based on minimum energy in (b,c).
  (d-c)~Quantum results for temperature-dependent Fourier amplitudes $A_n$ of torque computed via the orthogonalized finite-temperature Lanczos method (OFTLM) on 24 sites.
  $H\parallel b^\ast c$ shown in (b,d) and $H\parallel ab$ in (c,e).
 The vertical lines in (f) and (g) are guides to the eye for
estimates  of $T_{\rm N}$.
    }
  	\label{Fig4}
\end{figure}

{\it Field rotations in the honeycomb plane} [see Fig.~\ref{Fig2}(b))].
As the magnetic field is rotated in the honeycomb plane
a strong six-fold torque signal appears
below $T_\text{N}$ [see Figs.~\ref{Fig3}(b,d)], which we attribute to the ground state switching between different zigzag magnetic domains upon field rotation
(see Figs.~\ref{Fig4}(a)].
The torque signal becomes increasingly saw-tooth-like at 2~K
by increasing the applied field strength
[see Fig.~\ref{Fig2}(f)]
but no other
additional transitions are detected up to 16~T.
This is in contrast to $\alpha$-RuCl$_3$ where several qualitative changes occur due to in-plane field-induced transitions \cite{Froude2024}.  Note that
the torque signal above $T_\text{N}$ is much smaller in magnitude than inside the
zigzag phase (or compared with
the other orientation),
induced by extrinsic effects due to a small misalignment between the $c$-axis and the sample platform rotation axis, in addition to the sample weight (the $A_1$ component)
[see Figs.~\ref{Fig3}(d) and \ref{Fig2}(d)].
On the other hand, a residual two-fold periodicity
in $\alpha$-RuCl$_3$ was associated to strain within the crystal \cite{Froude2024}.

To parameterize the six-fold sawtooth shape of the torque signal
more precisely, we use a phenomenological 
form $ \tau(\phi) = A' \left(\frac{\gamma+1}{2}\right) \frac{\sin{6\phi}}{\sqrt{\cos^2{3\phi} + {\gamma^2}\sin^2{3\phi}}} $
where $A'$ is the torque amplitude and $\gamma$
characterises the sawtooth anisotropy ($\gamma=1$ for a pure sinusoid).
The temperature dependence of $A'$ acts like an order parameter for the paramagnetic to zigzag order transition [see Fig.~\ref{Fig3}(f)] whereas $\gamma$ varies between 1 to 2.4. The temperature dependence of $A'$ can be captured well by a phenomenological form $A'(T)=A'_{0}\left[1-(T/T_\text{N})^\alpha\right]^\beta$ with $T_\text{N}=35.4(3)~\mathrm{K}$, $\alpha=2.9(2)$ and $\beta=1.1(1)$ for sample S1
and similarly
for sample S2
[see Fig. S5 in the SM \cite{SM}].
Meanwhile,
muon spin rotation studies suggest that the
internal magnetic field
in $\alpha$-RuBr$_3$
characterizes a three-dimensional system.
Upon increasing magnetic field at 2~K,
$A'$ and $\gamma$ grow quadratically ,
but $\tau $ is significantly suppressed below 5~T, as shown in
Fig.~\ref{Fig2}(h)
[see also Fig.~S4 for S1 and Fig.~S6 for S2
in the SM \cite{SM}].
We attribute this effect to magnetic domains being pinned below a threshold applied in-plane field, $H_0$, not being able to switch between the three symmetry-equivalent domains as the magnetic field is rotated, resulting in a suppressed angular-dependence of the free energy and a much reduced torque signal.

{\it Theoretical calculations.}
We compare the experimental torque data to calculations in
extended Kitaev models
referred to as $JK\Gamma\Gamma'$-models.
 Here, we consider the nearest-neighbor couplings from the \textit{ab-initio} study of Ref.~\cite{Kaib2022}:
  $K_1\simeq -4$~meV for Kitaev exchange,  $J_1\simeq -2.9$~meV for Heisenberg and $\Gamma_1\simeq 2.8$~meV, $\Gamma'_1\simeq -0.5$~meV for symmetric off-diagonal exchanges with an easy-plane $g$-tensor ($g_{ab} = 2.32$, $g_c=1.88$). In order to capture the experimentally observed stability of the zigzag phase against field and temperature,
we adjust in our single-layer calculations an effective $J_3=1.5$~meV,
which accounts for larger longer-range interactions.
This results in a model for $\alpha$-RuBr$_3$ qualitatively similar
to the related compound $\alpha$-RuCl$_3$,
  except for the different longer-range interactions
\cite{Kaib2022,shen2024MagneticVsNonmagnetic}.
This is contrast to RuI$_3$,
which does not display any long-range order.
This different behavior has been attributed to
significantly modified nearest-neighbor interactions ($J_1$ and $\Gamma_1$) \cite{Kaib2022} as compared to  $\alpha$-RuCl$_3$, as well as important second- and third-neighbor anisotropic interactions, following recent
torque measurements~\cite{Ma2024}.

Figure~\ref{Fig4} shows the theoretical results of the described model for $\alpha$-RuBr$_3$, featuring a comparison between the classical zero-temperature results and finite-temperature observables using the orthogonalized finite-temperature Lanczos method (OFTLM) \cite{morita2020FinitetemperaturePropertiesKitaevHeisenberg} on a 24-site periodic cluster.
The zigzag order can occur in three symmetry-equivalent domains illustrated in Fig.~\ref{Fig4}(a). The energy of each domain is minimized when the applied field is parallel to the bond that contains antiferromagnetically opposed magnetic moment directions, [see Fig.~\ref{Fig4}(b,c)], e.g. for the domain with propagation vector $\mathbf Q=\mathrm{Y}$ the energy is minimized for $H\parallel b^\ast$. This yields for rotations in the $ab$-plane [Fig.~\ref{Fig4}(c)] the six-fold periodic torque signal 
with stable equilibrium positions 
when $H \parallel b^\ast$ and 
six-fold rotated directions, with discontinuities in $\tau$ at angles where the energetically favored domain switches, as shown in Fig.~\ref{Fig4}(e). These features are in good agreement with the experimental data of $\alpha$-RuBr$_3$ shown in Fig.~\ref{Fig2}(f) and were also
discussed in the case of $\alpha$-RuCl$_3$ \cite{Froude2024}.

On the other hand, when the field is rotated out of the honeycomb $ab$-plane towards $c$ [see Fig.~\ref{Fig4}(b)], all domains gain energy, which is a consequence of both the effective easy-$ab$-plane exchange anisotropy created by a positive $\Gamma_1$-exchange and the easy-plane $g$-tensor of the model, similar to $\alpha$-RuCl$_3$ \cite{Riedl2019}. Interestingly, upon rotating the magnetic field from $b^\ast$ to  $c$, the $\mathbf Q=\mathrm{Y}$ domain is slightly undercut in energy by another domain before all domains become degenerate by symmetry when $H\parallel c$.
This behaviour produces clear discontinuities in the classical calculation of $\tau$ shown in Fig.~\ref{Fig4}(d) \cite{footnoteOldTheory}.
The measured $\tau$ is steepest for $H\parallel c$ and has a shoulder-like feature for $\theta\approx 30^\circ,\, 210^\circ$ [see Fig.~\ref{Fig2}e], which could be a consequence of the domain switching effect.
Given that the energy differences between the three domains for field close to the $c$-axis are smaller than that
for field along different directions in the honeycomb $ab$ plane [see.~Fig.~\ref{Fig4}(b,c)],
each sample will likely stabilize different domain populations.
Fig.~\ref{Fig4}(b) shows how the energetically favorable domain changes with angle non-smoothly while rotating the field in the crystallographic $b^\ast\!c$ plane, giving the angular dependence of the torque a double-humped shape similar to experiment, as shown in Fig.~\ref{Fig4}(d).
While the employed $JK\Gamma\Gamma'$-model is 
two-fold symmetric around the $b^\ast$ axis, leading to antisymmetric torque around this axis, the shoulders seen in experiment do not follow this symmetry  [see Fig.~\ref{Fig2}(e)].
Indeed, the lack of this symmetry is consistent with the material's $R\bar{3}$ space group,
and could possibly be captured in models with interactions beyond $JK\Gamma\Gamma'$-exchanges, which go beyond the scope of this study. Note, that, for rotations in the $ac$ plane, no two-fold symmetry around $a$ is present in $JK\Gamma\Gamma'$-models, leading to a less symmetric torque for that rotation plane also in our employed model
[see Figs. S1(a,d,g) in SM \cite{SM}].

By performing finite-temperature quantum calculations,
we assess the temperature dependence of
the Fourier amplitudes of torque, as shown in Fig.~\ref{Fig4}(f,g).
We find
that $A_6$ increases
significantly inside the zigzag phase
whereas $A_2$ develops a broad peak due to thermal and quantum fluctuations,
in good agreement with the experimental
data in Fig.~\ref{Fig3}(c,d).
However, due to finite-size effects, the thermal phase transition are significantly washed out, as compared with experiments.

{\it Conclusion.}
We have studied in detail the magnetic anisotropy of the candidate Kitaev magnet $\alpha$-RuBr$_3$ using torque magnetometry and combined
with calculations considering the $JK\Gamma\Gamma'$ model.
Our studies clearly show
that $\alpha$-RuBr$_3$ has an easy-plane anisotropy
and the temperature dependence of the two-fold amplitude follows closely the powder-averaged magnetic susceptibility.
In-plane rotations exhibit a six-fold periodic saw-tooth torque up to 16~T
which reflect the symmetry of the lattice,
the domain formation and  the strength of
long-range interactions in this family of compounds.
$\alpha$-RuBr$_3$ is a model system
with a robust zigzag phase
due to the interplay of Kitaev
and significant long-range interactions.
Further studies
in very high magnetic fields
are required to suppress the
zigzag order in $\alpha$-RuBr$_3$
and stabilize other field-induced magnetic phases.

\paragraph{Acknowledgements.}
This work was partially supported by the EPSRC (EP/I004475/1),
the Oxford Centre for Applied Superconductivity
and the ISABEL project of the European Union’s Horizon's 2020 Research and Innovation Programme Grant Agreement Number No 871106. We acknowledge financial support from the Oxford University John Fell Fund.
AIC acknowledges an EPSRC Career Acceleration Fellowship (EP/I004475/1).
RC acknowledges support from the European Research Council under the European Union’s Horizon's 2020 Research and Innovation Programme Grant Agreement Number 788814 (EQFT).  DK and RV acknowledge support by the Deutsche Forschungsgemeinschaft (DFG, German Research Foundation) for funding through TRR 288 — 422213477 (project A05, B05).  RV, RC and AIC are grateful for hospitality by the Kavli Institute for Theoretical Physics (KITP) where part of this work was discussed,
supported in part by the National Science Foundation under Grants No. NSF PHY-1748958 and PHY-2309135.

\bibliography{RuBr3_bib}

\begin{thebibliography}{20}
\bibitem [{foo()}]{footnoteOldTheory}%
  \BibitemOpen
  \href@noop {} {}\bibinfo {note} {We note that this domain switching feature
  for field rotated in the $b^\ast\!c$-plane close to the $c$-axis was not
  detected in classical calculations of a comparable model for
  $\alpha$-RuCl$_3$ in Ref.~\cite{Riedl2019}, which only considered the
  energetics of the single $\mathbf Q=\mathrm{Y}$ domain.}\BibitemShut {Stop}%
\end{thebibliography}

\end{document}